\documentclass[sigconf]{acmart}
\usepackage{graphicx}
\usepackage{natbib}
\usepackage{booktabs}
\usepackage{caption}

\title{\textbf{Exploring the Advances in Using Machine Learning to Identify Technical Debt and Self-Admitted Technical Debt}}

\author{Eric L. Melin}
\affiliation{
  \institution{\textit{Computer Science Department}}
  \institution{\textit{Boise State University}}
  \country{Boise, ID, USA}
}
\email{ericmelin@u.boisestate.edu}

\author{Nasir U. Eisty}
\affiliation{%
  \institution{\textit{Computer Science Department}}
  \institution{\textit{Boise State University}}
  \country{Boise, ID, USA}
}
\email{nasireisty@boisestate.edu}

\setcopyright{acmcopyright}
\copyrightyear{2024}
\acmYear{2024}

\acmPrice{15.00}
\acmISBN{978-1-4503-XXXX-X/18/06}

\begin{document}

\label{abstract}
\begin{abstract}

\textbf{\textit{Context:} }
In software engineering, technical debt, signifying the compromise between short-term expediency and long-term maintainability, is being addressed by researchers through various machine learning approaches.
\textbf{\textit{Objective:} } 
This study seeks to provide a reflection on the current research landscape employing machine learning methods for detecting technical debt and self-admitted technical debt in software projects and compare the machine learning research about technical debt and self-admitted technical debt.
\textbf{\textit{Method:} }  
We performed a literature review of studies published up to 2024 that discuss technical debt and self-admitted technical debt identification using machine learning. 
\textbf{\textit{Results:} }
Our findings reveal the utilization of a diverse range of machine learning techniques, with BERT models proving significantly more effective than others.
\textbf{\textit{Conclusions:} }
This study demonstrates that although the performance of techniques has improved over the years, no universally adopted approach reigns supreme. The results suggest prioritizing BERT techniques over others in future works.

\end{abstract}


\maketitle

\label{introduction}
\section{Introduction}

Technical debt (TD), a metaphor initially introduced by Cunningham~\cite{cunningham_wycash_nodate}, represents the technical trade-offs that offer short-term benefits while possibly compromising the long-term integrity of a software system. 
Similarly, self-admitted technical debt (SATD), proposed by Potdar and Shihab ~\cite{potdar2014exploratory}, addresses intentionally introduced technical debt by developers, such as temporary workarounds.
Despite the concept of technical debt originating nearly three decades ago, it has only recently gained heightened attention from researchers in the past few years.

The decision to incur TD can expedite software development, enabling teams to meet tight deadlines or deliver functionality. 
Still, if left unaddressed, it can accumulate interest over time, akin to financial debt, compromising the quality and maintainability of a software system. 
TD is prevalent and regarded as a critical issue in the software industry, with an estimated global cost of 500 billion USD as of 2010 ~\cite{tom_exploration_2013}. 

TD and SATD identification, like many manual processes, would benefit massively from automation through a tool or service. Unfortunately, however, automating TD identification accurately is a great challenge and still requires much more work. The current state-of-the-art tools today are still far from perfect and are part of the reason for the influx of researchers coming to the intersection between technical debt and machine learning. Lenarduzzi et al.~\cite{lenarduzzi_towards_2019} reviewed SonarQube, a tool adopted by more than 98\% of public projects. Results from this study found 202 violations, of which only 26 were low fault-prone, and found that current instruments for SonarQube's estimations are not mature yet. 

However, the recent surge in research and application of machine learning (ML) techniques has led to an exciting intersection with TD/SATD identification in the research community, showing considerable promise. 
Due to the rapid advancements in ML techniques and the critical nature of managing TD/SATD in software systems, there have been instances of related work, such as a systematic mapping study on technical debt and its management~\cite{li_systematic_2015}, an exploration of ML across TDM activities~\cite{tsintzira2020applying}, and an exploration of ML approaches across all intelligent techniques~\cite{albuquerque_comprehending_2022}.

This paper presents a comprehensive, up-to-date reflection on ML techniques employed in research to identify TD and SATD. 
Our reflection paper distinguishes from prior literature reviews through the following points of differentiation:
\begin{itemize}
\item Sierra et al.~\cite{sierra2019survey} performed a literature survey classifying existing SATD work into detection, comprehension, and repayment and did not focus on the comparison of different used techniques. This work comprises only studies between 2014-2018, did not report following the procedures outlined by Kitchenham and Charters~\cite{keele2007guidelines}, and did not report consistent evaluation metrics on the techniques they covered.

\item Sutoyo and Capiluppi~\cite{sutoyo2023detecting} performed a systematic literature review on automated approaches to detect SATD using NLP + varying combinations and grouped all approaches into NLP groups. In this work, they focused on SATD identification and categorization performance, as well as SATD detection in various software development activities. This work consists of only studies between 2002 and 2022, focuses on studies using NLP to detect technical debt, and does not use Google Scholar as a source. Furthermore, it is imperative to note the contrasting length between their extensive 62-page review and our concise 7-page reflection.

\end{itemize}

Given the swift progress in ML techniques, the heightened focus on TD/SATD in recent years, differences in prior works, and the time that has elapsed since the publication of these related works, there is a compelling demand for updated research and methodologies in this domain specifically to discover which ML techniques are most effective for identifying TD/SATD.

To overcome these limitations, we conducted a comprehensive literature review to examine past and current research on the application of ML techniques for identifying both TD and SATD. 
We employed a multistep approach to gather the most relevant papers using ACM digital library, Google Scholar, IEEE Xplore, Scopus, and Springer for querying and paper snowballing. 
The key contributions and main findings of this study can be summarized as follows:
\begin{itemize}
    \item \textit{Identifying current TD and SATD identification methods.}
    \item \textit{Identifying the most effective ML approaches for TD and SATD identification.}
\end{itemize}

\label{methodology}
\section{Methodology}
This section delves into our research methodology and the research questions guiding our study.

\subsection{Search Strategy}

\begin{figure}
    \centering
    \includegraphics[width=1\linewidth]{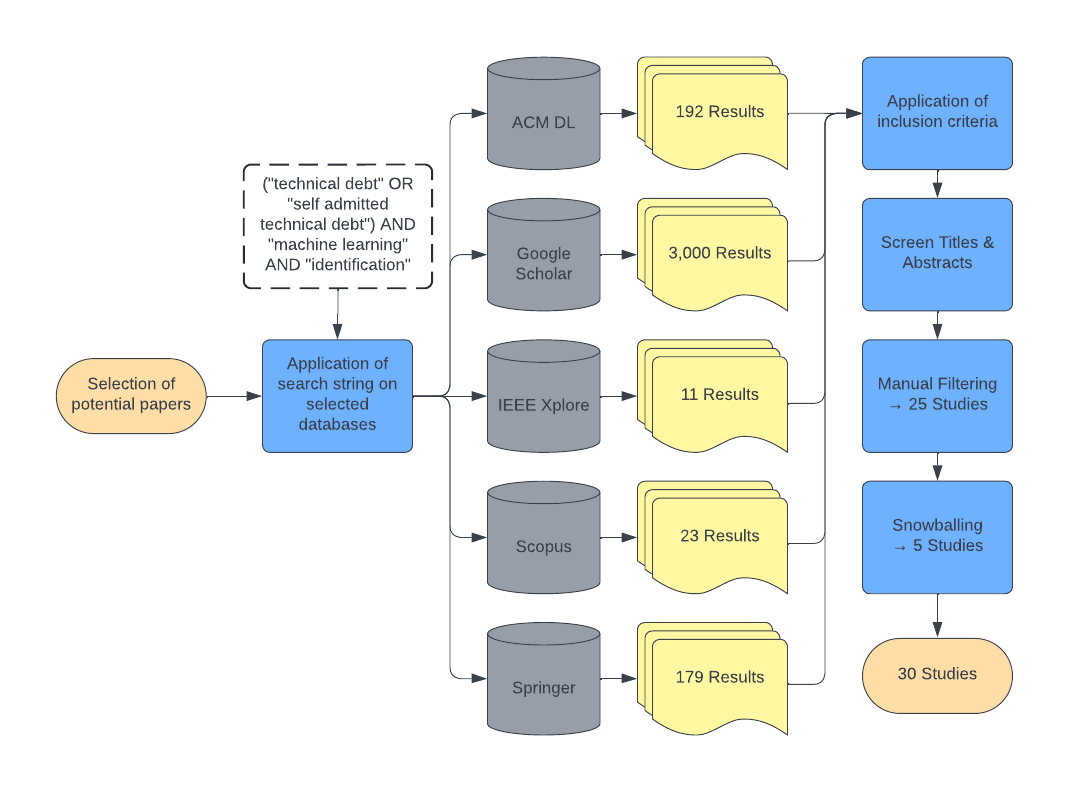}
    \caption{Paper Selection Process}
    \label{fig:Paper Selection Process}
\end{figure}

\begin{table}
    \centering
    \begin{tabular}{|p{0.18\linewidth}|p{0.72\linewidth}|}
        \hline
        \textbf{Inclusion Criteria} & - The article covers at least one of our research questions partially or fully. \newline - The article is written in English. \newline - The article is published in a scholarly journal or conference/workshop/symposium proceedings. \newline - Any article published after 2015. \\
        \hline
        \hline
        \textbf{Search String} & ("technical debt" OR "self admitted technical debt") AND "machine learning" AND "identification" \\
        \hline
    \end{tabular}
    \caption{Protocol Summary}
    \label{tab:protocol_summary}
\end{table}


To undertake this analysis, we adopted a multistep approach to select pertinent papers to form the basis of our study, as seen in Figure~\ref{fig:Paper Selection Process}. Our initial phase involved an extensive search on 5 varying sources, using a predefined search string that aligned with our research questions, as shown in Table~\ref{tab:protocol_summary}.  This initial search established a foundation for reviewing studies.


To ensure the relevance and quality of the studies included in this review, clear inclusion criteria (refer to Table~\ref{tab:protocol_summary}) were established. The process of study selection, as depicted in Figure~\ref{fig:Paper Selection Process}, commenced with our initial search on 5 sources, which yielded a total of 3,405 studies, including duplicates. Notably, papers published before 2016 were excluded due to the evolving landscape in the field and the lack of existing literature at the time. Papers published in a language other than English were omitted to maintain consistency and accessibility. Subsequently, to refine the search further, titles and abstracts were screened according to predefined keywords (refer to Table~\ref{tab:protocol_summary}) and relevance to the predefined research questions (RQs). This manual screening process resulted in the identification of 25 studies meeting the specified criteria.

Building upon the outcome of the database searches, we applied forward and backward paper snowballing techniques. By tracing the citations and references of identified papers, we aimed to widen our search and discover more studies that might have eluded our initial search. This iterative process not only broadened the scope of our search but also resulted in the inclusion of 5 tightly related papers.



\subsection{Data Collection and Analysis}
To perform data collection from the collected papers, we followed the structure of data collection forms from Kitchenham and Charters~\cite{keele2007guidelines} to collect all the information needed to address the review questions. The forms included data as (i) title, (ii) date of data extraction, (iii) authors, (iv) journal, (v) publication details, (vi) year of publication, (vii) TD or SATD, (viii) ML approaches used, (ix) methodology, (x) results, (xi) future work, (xii) limitations, (xiii) data analysis, and (xiv) space for additional notes. 
The first author performed this data extraction independently.

After the data extraction, we synthesized the results of the primary studies according to Kitchenham and Charters~\cite{keele2007guidelines}. The descriptive synthesis of the data can be found in the results sections and a quantitative tabular form in Table~\ref{tab:comparison} based on the outcome of the primary study: did they cover TD or SATD, what ML approaches did they use, results, and objective of the paper. We also highlighted best-performing techniques when multiple were used. The studies are listed in Table~\ref{tab:comparison} chronologically as they are found in this review.

\subsection{Research Questions}
The goal of this study is to conduct a thorough analysis of existing machine learning literature, with a specific focus on comprehending contemporary approaches utilized for managing TD and SATD. The study aims to explore these ML approaches in regard to three key aspects: (a) the current machine learning strategies, (b) the current machine learning solutions, and (c) the mapping between TD and SATD, specifically in the context of leveraging ML for TD management in software. Therefore, we pose the following research questions.

\begin{enumerate}
    \item[\textbf{RQ1:}] \textbf{\textit{Which methods are employed for identifying TD?}}
    
By finding the best approaches, future researchers can then focus on improving the best-known approach or attempt to take a novel approach not yet attempted. 
    
    \item[\textbf{RQ2:}] \textbf{\textit{Which methods are employed for identifying SATD?}}
    
    Similar to RQ1, by finding the best approaches, future researchers can then focus on improving the best-known approach or attempt to take a novel approach not yet attempted.
    
    \item[\textbf{RQ3:}] \textbf{\textit{How do current ML approaches compare between TD and SATD identification?}}
    
    By identifying the best identification techniques from both TD and SATD identification literature, we can see the differences in approaches that researchers have taken for each one.  By illuminating effective strategies from one, researchers can then attempt to reapply those strategies for the other.
\end{enumerate}


\label{results}
\section{Results}
This section presents the findings obtained from the studies retrieved in this literature review. The results are grounded on the set of 30 studies retrieved using the aforementioned search strategy. 

\subsection{RQ1: Which methods are employed for identifying TD?}
  In this section, we present our exploration of our search to offer a nuanced understanding of the diverse methodologies adopted by previous researchers.  Each study, a unique perspective on the intersection between ML and TD identification, provides its own meaningful insights.  Through these studies, our goal is to illuminate trends, inform, and contribute to the discourse of this exciting intersection between machine learning and software.

The analysis of the retrieved studies resulting from our search strategy revealed several ML approaches and statistical data across several studies being used for TD identification. Our sampled studies support the idea that while no single ML approach dominates the rest in the realm of TD identification, there are approaches that perform much better than others. 
Tsoukalas et al.~\cite{tsoukalas_machine_2021} tested the effectiveness of 7 ML classifiers, resulting in rankings and a benchmark for allowing ML to classify modules as high TD or not. They found that random  forest (RF), eXtreme gradient boosting (XGB), logistic regression (LR), and support vector machine (SVM) approaches are best for identifying TD.

Aversano et al.~\cite{aversano2023forecasting} employed multiple ML techniques to evaluate whether the quality metrics of a software system can be useful for the correct prediction of the technical debt. In the study, they employed multiple ML algorithms  including multiple linear regression (MLR), decision tree (DT), bagged decision tree (BDT), and RF.  After evaluation using root-mean-square error (RMSE) they found that RF yields the best performance.

Similarly, 2 papers discussed the effectiveness of ML approaches on code smell detection. Like the previous study~\cite{tsoukalas_machine_2021}, Cruz et al. ~\cite{cruz_detecting_2020} found RF and XGB to be the very best of the 7 ML techniques they compared, suggesting that these two techniques should be the pinnacle of future of tools and research regarding TD identification. Fontana et al.~\cite{arcelli_fontana_comparing_2016} found that across several code smells, the J48 family of algorithms and RF are the best classifiers of each tested code smell.  Furthermore, from their testing, they found that there appears to be no clear benefit in using boosting algorithms for detecting code smells.

Finally, Dai and Kruchten ~\cite{dai_detecting_2017} explored detecting TD through issue trackers, an avenue not discussed by the previous studies.  The authors applied natural language processing (NLP) and Naive Bayes (NB) classification to successfully identify common words that can be used as indicators of TD in issue trackers, showing that there is also great potential in TD identification through issue trackers.  

\begin{table*}
    \centering
    \caption{Comparison of ML Methods for TD/SATD Identification}
    \label{tab:comparison}
    \begin{tabular}{p{1.3cm}p{1.2cm}p{3.3cm}p{3.8cm}p{4.7cm}}
        \toprule
         \textbf{Ref. No.} & \textbf{TD/SATD} & \textbf{Tested Method(s)/Tools} & \textbf{Best Result(s)} & \textbf{Objective} \\
        \midrule
        \cite{tsoukalas_machine_2021} & TD & \textbf{RF}, \textbf{XGB}, LR, SVM, kNN, NB, DT & (F2) RF: 0.790, XGB: 0.788 & Classifying modules as high TD or not. \\
        \midrule
        \cite{aversano2023forecasting} & TD & MLR, \textbf{RF}, BDT, DT & RF & TD Forecasting. \\ 
        \midrule
        \cite{cruz_detecting_2020} & TD & NB, LR, NLP, DT, kNN, \textbf{RF}, \textbf{XGB} & (F1) RF: 0.861, XGB: 0.859 & Code smell detection. \\
        \midrule
        \cite{arcelli_fontana_comparing_2016} & TD & \textbf{J48 families}, JRIP, NB, \textbf{RF}, SVM & J48 family of algorithms, RF & Code smell detection. \\
        \midrule
        \cite{dai_detecting_2017} & TD & NLP & (F1) NLP: 0.76 & TD detection through issue trackers. \\
        \midrule
        \cite{sridharan_data_2021} & SATD & SMOTE, RF, \textbf{XGB}, LR & (F1) XGB: Within-Project 0.755, Cross-Project 0.729 & Effect of data balancing on SATD detection. \\
        \midrule
        \cite{tu_debtfree_2022} & SATD & \textbf{DebtFree(0)}, \textbf{CNN}, JitterBug, DebtFree(100) & (Median F1) CNN: 61, DebtFree(0): 57 & Proposed the DebtFree framework for identifying ``easy'' and ``hard'' SATD. \\
        \midrule
        \cite{zampetti_recommending_2017} & SATD & DT, NB, \textbf{RF}, RT, Bagging w/ DT & (F1) RF: Within-Project 0.4715 & Developed TEDIOUS for detecting SATD. \\
          \midrule
        \cite{yin_two-stage_2023} & SATD & DT & (F1) DT: Within-Project 0.784, Cross-Project 0.719 & Proposed a two-stage approach to identify and interpret SATD. \\
        \midrule
        \cite{huang_identifying_2018} & SATD & NB & (F1) NB: 0.737 & Proposed an approach to detect SATD using text mining. \\
        \midrule
        \cite{alhefdhi2022framework} & SATD & \textbf{MNB}, SMV, RF, RNN, CNN, \textbf{LSTM} & (F1) MNB: 0.251, LSTM: 0.331 & A framework for technical debt identification and description. \\
        \midrule
        \cite{flisar_enhanced_2018} & SATD & word2vec & (Precision) word2vec: 82\% & Exploration of training word embeddings to detect SATD. \\
        \midrule
        \cite{santos_long_2020} & SATD & word2vec w/ LSTM & (F1) word2vec w/ LSTM: design SATD 0.670, requirement SATD 0.606 & Exploration of using a LSTM with word2vec to identify requirement SATD and design SATD. \\
        \midrule
        \cite{wattanakriengkrai_identifying_2018} & SATD & N-gram IDF w/ autosklearn & (F1) N-gram IDF w/ autosklearn: design SATD 0.640, requirement SATD 0.637. & Proposed a model for identifying requirement SATD and design SATD. \\
        \midrule
         \cite{maipradit2020automated} & SATD & N-gram IDF w/ autosklearn & (F1) N-gram IDF w/ autosklearn: 0.73 & Proposed a tool to identify ``on-hold'' SATD. \\
        \midrule
        \cite{maipradit_wait_2020} & SATD & N-gram TF-IDF w/ autosklearn & (F1) N-gram TF-IDF w/ autosklearn: 0.77 & Proposed a tool to identify ``on-hold'' SATD. \\
        \midrule
        \cite{ren_neural_2019} & SATD & CNN & CNN: Within-Project 0.752, Cross-Project 0.766 & Proposed a CNN approach for detecting SATD.  \\
        \midrule
        \cite{zampetti_automatically_2020} & SATD & CNN \& RNN & (F1) CNN \& RNN: 55.82\% & Classifier for identifying and recommending 6 SATD removal strategies. \\
        \midrule
        \cite{li_identifying_2022} & SATD & SVM, \textbf{NBM}, kNN, \textbf{LR}, RF, Text GCN, \textbf{Text CNN} & (F1) Text CNN: 0.597, NBM: 0.529, LR: 0.515 & Proposed an approach for identifying SATD in issue tracking systems. \\
        \midrule
        \cite{li_automatic_2023} & SATD & MT-Text-CNN & (F1) MT-Text-CNN: 0.611 & Proposed an approach for identifying SATD by combining multiple sources. \\
        \midrule
        \cite{maldonado_using_2017} & SATD & NLP & (F1) NLP: design SATD 0.620, requirement SATD 0.403 & Presented an approach to identify design SATD and requirement SATD. \\
        \midrule
        \cite{rantala_predicting_2020} & SATD & NLP & (AUC) NLP: 0.7411 & Proposed an approach for detecting SATD through commit contents. \\
        \midrule
        \cite{sala2021debthunter} & SATD & NLP & (F1) NLP: 0.750 & Proposed DebtHunter, an approach for identifying and classifying SATD. \\
        \midrule
        \cite{sharma_self-admitted_2022} & SATD & ME, SVM, LR, CNN, \textbf{ALBERT}, \textbf{RoBERTa} & (F1) ALBERT: 0.8621, RoBERTa: 0.8609 & Exploration of using PTMs to detect SATD. \\ 
        \midrule
        \cite{aiken_measuring_2023} & SATD & BERT & (F1) BERT: 0.868 & Proposed a novel BERT approach for SATD detection. \\
        \bottomrule
    \end{tabular}
\end{table*}

\subsection{RQ2: Which methods are employed for identifying SATD?}

Similar to the studies reviewed for TD identification, researchers investigating SATD identification have been using a wide variety of ML approaches to identify SATD, and there is no universal state-of-the-art approach.  
There has been much work with RF, LR, SVM, NLP, and NB. Unlike the study pool we accumulated for TD identification, this study pool contained several studies exploring the use of text mining,  bidirectional encoder representations from transformers (BERT), and convolutional neural networks (CNN).

\subsubsection{\textbf{Data Management}}
To begin, Sridharan et al. ~\cite{sridharan_data_2021} found that data balancing improves SATD identification in within-project and cross-project setups. Their results showed that the data level balancing techniques Synthetic Minority Over-
sampling Technique (SMOTE), RF, and XGB are all reasonable approaches for maximizing precision, recall, F1, or AUC-ROC, depending on the goal. Their results showed that no single balancing technique provides higher performance across multiple metrics, showing that no one technique dominates the others.

Tu and Menzies~\cite{tu_debtfree_2022} proposed DebtFree, a framework to minimize labeling cost in SATD identification using semi-supervised learning (SSL). Their SSL approach involves training models on a combination of labeled and unlabeled data, which is most useful when labeled data is scarce or expensive to obtain. The framework then utilizes unsupervised learning for pseudo-labeling and filtering and then active learning for iterative model training and updating. The authors found that when comparing against the then state-of-the-art semi-supervised learning approach Jitterbug ~\cite{yu_identifying_2022}, and supervised learning approach ~\cite{ren_neural_2019}, DebtFree reduces the labeling cost and performs just as well without having access to the training data's labels. 

\subsubsection{\textbf{Trees}}
Zampetti et al. ~\cite{zampetti_recommending_2017} studied how 5 machine learning techniques (DT, Bayesian classifiers, RF, random trees (RT), and Bagging with DT) performed when tasked with recommending developers to admit design TD. Their results showed that RF outperformed all the techniques for within-project prediction with an average F1 score of 0.4715 without data balancing and 0.3604 when data balancing with SMOTE.  

Yin et al.~\cite{yin_two-stage_2023} proposed a two-stage approach to identify SATD using interpretable methods. In stage one, they combined a decision tree model with an integrated model to improve identification accuracy. They then use SHAP, LIME, and Anchors for feature extraction, interpretation, and analysis. The approach, on average, resulted in a within-project F1-score of 0.784 and a cross-project F1-score of 0.719.

\subsubsection{\textbf{Naive Bayes}}
Huang et al.~\cite{huang_identifying_2018} developed an ML model relying on several sub-classifiers. Each sub-classifier used the Naive Bayes (NB) multinomial, with information gain as feature selection. The average F1-score achieved was 0.737, proving to be a significant improvement over the baseline approaches in 2018.

Alhefdhi et al. ~\cite{alhefdhi2022framework} proposed a framework Self-Admitted Technical Debt Identification and Description (SATDID) for determining if technical debt should be self-admitted for an input code fragment. In the study they test Multinomial Naive Bayes (MNB), SVM, RF, Recurrent Neural Network (RNN), CNN. and Long Short-Term Memory neural (LSTM) network. They report that from the traditional ML algorithms, MNB performed the best with an F1-score of 0.251 and that the LSTM performed the best with an F1-score of 0.311

\subsubsection{\textbf{Word Embeddings}}
Flisar and Podgorelec~\cite{flisar_enhanced_2018} noticed that most NLP methods only used manually annotated data to train their classifiers. In their study, they used a large corpus of unlabeled code comments extracted from open-source projects to train word embeddings to detect SATD. In their approach, they train the word2vec model and report an 82\% correct prediction accuracy.

Santos et al. ~\cite{santos_long_2020} also used the word2vec word embeddings model but combined it with a LSTM neural network. In their work, they compared their model against other models for design and requirement SATD classification. To these ends, their model resulted in an average F1-score for design SATD classification of 0.670 and requirement SATD classification of 0.606. From these results, they found that it outperforms LSTM, auto-sklearn, and maximum entropy approaches.

\subsubsection{\textbf{N-Gram}}
Wattanakriengkrai et al.~\cite{wattanakriengkrai_identifying_2018}, similar to Santos et al. ~\cite{santos_long_2020}, focused on 2 types of SATD specifically design and requirement debt, but instead of using word embeddings used N-gram IDF and auto-sklearn for automated SATD identification. Their study resulted in an average F1-score for design debt of 0.640 and 0.637 for requirement debt.

Maipradit et al.~\cite{maipradit2020automated} presented an approach using N-gram IDF capable of detecting issues referenced in code comments, and to automatically detect instances of ``On-hold'' SATD, where the developer's comments indicate that a developer is holding off on future work due to a future event. Using n-gram + autosklean they achieved an F1-score of 0.73.

Maipradit et al.~\cite{maipradit_wait_2020} second approach used N-gram TF-IDF and auto-sklearn to identify cases of ``On-hold'' SATD. Their classifier achieved a mean F1-score of 0.77 and found that compared to Naive and TF-IDF baselines, N-gram TF-IDF outperforms them both on all measures. With their work, the authors confirm that their approach is positive for identifying on-hold SATD but did not consider other TD types in their work.

\subsubsection{\textbf{Nueral Networks}}
Ren et al.~\cite{ren_neural_2019} authored the earliest instance of SATD identification using a CNN found in our study. Here, the authors presented a novel CNN-based approach to code comments, which achieved a substantial improvement over previous text-mining approaches in both within-project and cross-project settings at the time. This improvement can be attributed to CNN's capability to learn to extract more meaningful and comprehensive SATD-indicating features than text-mining methods. This novel CNN, on average, resulted in a within-project F1-score of 0.752 and a cross-project F1-score of 0.766.

Zampetti et al.~\cite{zampetti_automatically_2020} built a multi-level classifier capable of identifying and recommending six SATD removal strategies: changing API calls, conditionals, method signatures, exception handling, return statements, or telling that a more complex change is needed. To build their classifier, they used a CNN and RNN and found that their evaluation is capable of predicting the type of change with an average F1-score of 55.82\% and AUC of 0.73.

In contrast to other methods mentioned, Li et al.~\cite{li_identifying_2022} noticed that in addition to source code comments, issue tracking systems have shown to be another rich source of SATD. They observed that no previous approaches specifically target SATD in issue tracking systems and proposed an approach for automatically identifying SATD in these systems by testing a variety of ML techniques consisting of SVM, NBM, k-Nearest Neighbor (kNN), LR, RF, Text Graph Convolutional Network (Text GCN), and Text Convolutional Neural Network (Text CNN). From their testings, they found that a Text CNN using random word embeddings achieves the highest F1-score of 0.597, only closely followed by NBM (0.529) and LR (0.515).    

Building upon their previous work~\cite{li_identifying_2022}, Li et al.~\cite{li_automatic_2023} found that there was a lack in current works for approaches designed to identify SATD from multiple sources such as commit messages, pull requests, source code comments, and issue tracking systems. Li et al.~\cite{li_automatic_2023} proposed a multitask text convolutional neural network (MT-Text-CNN) to fill this gap. Their approach outperforms baseline approaches, achieving an average F1-score of 0.611 when detecting 4 types of SATD from the 4 aforementioned sources. The findings of this paper also indicate that SATD is evenly spread among all sources and that issues and pull requests are the two most similar sources regarding the number of shared SATD keywords followed by commit messages and then followed by code comments.

\subsubsection{\textbf{NLP}}

Rantala and Mantyl~\cite{rantala_predicting_2020} also explored using NLP to detect SATD, but from code commit contents. In their work, they reproduce and improve on a prior work by Yan et al.~\cite{yan_automating_2019}. In the improvement study, the authors use multiple NLP methods, including bag-of-words, topic modeling, and word embedding vectors. Of these 3 techniques, the authors found that bag-of-words strongly outperforms the other two techniques, reaching a median AUC of 0.7411.

Sala et al. proposed DebtHunter, with a two-step classification phase: SATD identification and SATD classification into specific debt types. Using NLP, DebtHunter achieves an F1-score of 0.750 on test.

\subsubsection{\textbf{BERT}}
Very recently, researchers have also attempted to identify SATD with Bidirectional Encoder Representation from Transformers (BERT) models.  Sharma et al.~\cite{sharma_self-admitted_2022} pioneered using Pre-Trained Language Models (PTM) for SATD detection and performed a study on SATD identification in the R programming language.  They compared several classifiers, including Max Entropy (ME), SVM, LR, CNN, ALBERT, and RoBERTa.  They found that across average F1-scores the two PTMs ALBERT and RoBERTa outperformed all the models with average F1-scores of 0.8621 and 0.8609, respectively. The CNN followed with 0.7989, closely followed by ME at 0.7636. SVM and LR performed well below the other methods with average F1-scores of 0.6722 and 0.6637, respectively. In addition to pioneering the use of PTMs in SATD detection, they also pioneered studying the effect of lemmatization in ME. The findings were that lemmatization displays positive results in most SATD types provided a large dataset, but despite the improvement, it does not match the impressive performance of the PTMs.

Aiken et ak.~\cite{aiken_measuring_2023} also leveraged a BERT model for the detection of SATD. In their study, they compared their model with previous deep learning methods and applied stratified 10-fold cross-validation to report reliable F1-scores. They examined their model in both cross-project and intra-project contexts and found that their BERT model improves the best previous methods in 19 of the 20 projects in cross-project scenarios. Their best approach for cross-project on two datasets resulted in an average F1-score of 0.858 and 0.868. As far as the intra-project scenario goes, they found the lack of data present to overfit the results even when using data augmentation techniques and found that existing methods performed better.

\subsection{RQ3: How do current ML approaches compare between TD and SATD identification?}
A synthesis of our findings is present in Table~\ref{tab:comparison}, organized in the order of paper references in RQ1 and RQ2. The table rows are grouped based on similar methods for easier comparison.

Our exploration in RQ1 and RQ2 revealed that while there is no definitive state-of-the-art solution for either TD identification or SATD identification, certain methods outperform others.

From RQ1, it became evident that among the ML approaches employed in the studies, random forest and extreme gradient boosting yielded superior results. These studies also highlighted a correlation between code smell and TD, suggesting a propensity for TD in the presence of code smells. Notably, the absence of research utilizing CNNs and language models such as BERT for TD identification was surprising, indicating a promising avenue for future research.

RQ2 demonstrated that among the studies surveyed, a much wider variety of techniques are being leveraged for SATD identification. RQ2 shows that BERT models consistently outperformed all other ML techniques, achieving the highest F1-scores, closely followed by CNNs.


In contrast to our study, prior research lacks a comparison between TD and SATD identification techniques. The discrepancy in the quantity of papers retrieved between TD and SATD shown in Table~\ref{tab:comparison} underscores the difference in works between TD and SATD research.


\label{threats to validity}
\section{Threats to Validity}


\subsection{Internal Validity}
The internal validity of this study may be impacted by several factors. Firstly, the paper selection process heavily relied on the relevance identified through five databases: Google Scholar, IEEE Xplore, Scopus, Springer, and ACM Digital Library, using specific keyword combinations. While efforts were made to ensure comprehensive coverage, there remains a possibility of missing relevant studies that were not indexed or did not surface in our search results. Additionally, the process of paper snowballing, while employed to broaden the search, might have also introduced biases or oversights in the selection of studies. Furthermore, the inclusion or exclusion of papers was determined by a predefined search string and manual analysis, which could introduce subjectivity and potential errors.

\subsection{External Validity}
Despite the use of five diverse databases, the specific keyword search string and search strategies employed may have restricted the breadth of our search, potentially omitting relevant studies published in non-indexed sources or in languages other than English.

\subsection{Construct Validity}
In our study, construct validity may be influenced by the selection of the search string used for querying across the five databases. While efforts were made to select a comprehensive and representative search string, there is a possibility that some relevant terms were not included, leading to a partial representation of the domain. Furthermore, the manual content analysis conducted during the paper selection process may introduce subjectivity and biases, potentially affecting the accuracy and completeness of the included studies.

\label{conclusion}
\section{Conclusion}

In this paper, we analyzed the existing literature related to using ML techniques for TD and SATD identification using a multiple-step approach to gather literature in order to understand: 1) existing ML techniques used for TD identification; 2) existing ML techniques used for SATD identification; 3) the differences in the research between using ML to identify TD and SATD.

The results present the ML techniques used for both TD and SATD identification and the effectiveness of said techniques in order to outline the future research trajectory on the use of ML for both identifying TD and SATD. Moving forward, researchers should consider prioritizing the best approaches from this literature review, consider the distinctions between TD and SATD, and continue to attempt to improve the performance of TD/SATD identification tools in order to future improve software quality.   

\bibliographystyle{ACM-Reference-Format} 
\bibliography{sec_bibliography}


\end{document}